# FR TRUST: A Fuzzy Reputation Based Model for Trust Management in Semantic P2P Grids

Saeed Javanmardi[a], Mohammad Shojafar[2,*], Shahdad Shariatmadari[3] and Sima S. Ahrabi[4]

[1] *Department of Computer Engineering, Dezful branch, Islamic Azad University, Dezful, Iran*
[2] *Dept. of Information Engineering, Electronic and Telecommunication (DIET), Sapienza University of Rome, Rome, Italy Email: shojafar@diet.uniroma1.it*
[3] *Department of Computer Science and Information Technology, University Putra Malaysia, Serdang, Malaysia*
[4] *Department of Mathematic, Faculty of Science, University Technology Malaysia, 81310 UTM Johor Bahru, Johor Darul Takzim, Malaysia*

**Abstract.** Grid and peer-to-peer (P2P) networks are two ideal technologies for file sharing. A P2P grid is a special case of grid networks in which P2P communications are used for communication between nodes and trust management. Use of this technology allows creation of a network with greater distribution and scalability. Semantic grids have appeared as an expansion of grid networks in which rich resource metadata are revealed and clearly handled. In a semantic P2P grid, nodes are clustered into different groups based on the semantic similarities between their services. This paper proposes a reputation model for trust management in a semantic P2P Grid. We use fuzzy theory, in a trust overlay network named FR TRUST that models the network structure and the storage of reputation information. In fact we present a reputation collection and computation system for semantic P2P Grids. The system uses fuzzy theory to compute a peer trust level, which can be either: Low, Medium, or High. Our experimental results demonstrate that FR TRUST combines low (and therefore desirable) a good computational complexity with high ranking accuracy.

## 1. Introduction

P2P grid computing combines and integrates grid and P2P technologies to implement peer-to-peer communications with greater distribution and scalability [1]. Trust management is a complicated and difficult task in such an environment, because resources are geographically distributed and belong to distinct organizations [2]. In a P2P environment, some peers may provide services with low quality and may not promise to satisfy user requirements.

An unfavorable situation arises when providers offer incorrect information about their resources/services to exaggerate the quality of their services [3]. To inspire resource sharing among nodes and protect against malicious node behaviors, a reputation system for trust management is necessary. Such system allows nodes to estimate the trustworthiness of others and to selectively interact with the more respectable ones and avoid egocentric, dishonest, and malevolent node behaviors [4, 5]. This feature explains in several recent works such as [6-8]. In this paper, we combine reputation-based trust, fuzzy theory, and a web of trust and recursively propagating trust [9, 10] to propose a model for trust management in semantic P2P grids.

Our approach is usable in both P2P Grid systems and semantic P2P grid systems. In fact the usable of semantic is just to create Semantic Overlay Network (SON) which has lots of advantages for clustering the network. SON improves query performance and maintains a high degree of node autonomy. In another word, semantic is just used for clustering the grid environment based on the similarity between re-sources while fuzzy theory is used for trust management.

According to recursively propagating trust, if A has trust x in B and B has trust y in C, then A must have some trust z in C which is a function of x and y. A web of trust is based on recursively propagating trust. The coordinator is trusted by all nodes in the group and calculates the trust level of each node based on the information received from the other nodes. (How the coordinator of each group is elected lies beyond the scope of this paper.) Each node maintains reputation information about other nodes, thus generating the web of trust [11].

In this paper we propose a fuzzy reputation-based model for trust management in a semantic P2P grid network. In Semantic P2P environments, nodes are clustered based on their interests and similarities; thus in a semantic P2P grid network, nodes can be clustered based on the semantic similarity of their resources [12]. In our model, we use agents which are responsible for trust management. Our model calculates reputation scores by aggregating feedback from nodes to determine nodes' trustworthiness.

The rest of this paper is organized as follows. Section 2 provides preliminaries: we take a brief look at trust, reputation and fuzzy logic. Section 3 presents related work. Section 4 discusses our proposed model, and Section 5 presents its performance evaluation and experimental results. Finally, Section 6 presents our conclusions and future researches.

## 2. Preliminaries

This section provides an introduction to trust, reputation and fuzzy theory, as a background for our model.

### 2.1. Trust

Trust serves as the foundation for both human society and cyberspace security. Each one of us is aware of the significance of trusting somebody. The nature of trust is usually decentralized, since the parameters of trust are usually individual. Trust can be described as an individual's certainty that a certain party will exhibit an anticipated behavior despite guarding or managing the individual. Trust is most commonly measured practically and produces a good effect in vague modifying conditions. Trust is not a black-and-white commodity. Frequently, there is a grey area in a computer site's quality of reliability [13, 14]. As with human associations, trust in a computational context can be captured well by a linguistic name in a numerical fashion. The concept behind trust is a composite connected to a solid confidence in the applicability of adjectives, for example, the trustworthiness, truthfulness and capability of the trusted thing. Authors in [15] explain trust as follows: trust is the firm belief in the competence of an entity to act as expected, where the firm belief is not a fixed value associated with the entity but is rather a function of the entity's behavior and applies only within a specific context at a given time.

### 2.2. Reputation

Reputation is determined as an estimate of reliability in the sense of trustworthiness. Reputation systems [16] offer a basis for developing trust through social control without trusting third parties, by way of community-based responses about past experience of entities. This helps in reaching recommendations and opinions about the quality and persistence of transactions [17, 18]. Reference [15] explain the reputation of an entity as the anticipation of its acting in a manner that is dependent upon other entities' supervisions or upon knowledge about the entity's past manner of acting.

### 2.3. Fuzzy theory

Fuzzy theory that is capable of handling several types of ambiguity [19]. Where x is a fuzzy set and u is a related object, the statement "u is a member of A" is not always either exactly true or exactly false. It may be true only to some degree, the degree to which u is in fact a member of x. A crisp set is specified in such a way as to divide everything under discussion into two groups: members and non-members. A fuzzy set can be specified in a mathematical form by assigning to each individual in the universe of discourse a value giving its degree of membership in the fuzzy set [20].

## 3. Related Work

Trust management has recently become a very practical and powerful tool in some special environments where a lack of previous knowledge about the system can guide participants to unwanted conditions, particularly in virtual environments where users do not know each other. This section presents some of the most typical trust models for distributed environments.

Kamvar et al. [21] proposed a trust model which is described by the assignment of a unique global trust value to each node in a P2P environment, according to the node's history of actions. This trust model assumes that some pre-trusted nodes are available and that they are trusted by all nodes in the environment. Nodes carry out a distributed computation coming closer to the eigenvector of the trust matrix over the nodes.

A trust model proposed by Xiong et al. [22] combines several significant features related to reputation-based trust management in distributed systems, such as: the feedback a node receives from other nodes, the total number of a node's transactions, and the dependability of the endorsements given by a node.

A trust model proposed by Zhou et al. [23] implements power-law response attributes that have been found suitable in dynamic P2P environments, either structured or unstructured. In this model, only a small number of power nodes that are most respectable, based on the power-law response attributes, are selected in a dynamic manner, using a ranking mechanism.

Karaoglanoglou and Karatza in [24] proposed a trust-aware resource discovery model that uses good trustworthiness values to promise gratification of needs in a grid environment. It is not obvious how trust is computed for each virtual organization (VO) in the grid environment. The authors only surmise random values for all VOs.

Pooranian et al. in several recent works [25-29] focused mainly on Resource discovery models respecting makespan of independent workloads in heterogeneous VOs based on trusted models and securities by applying various hybrid soft computing methods such as Particle Swarm Optimization (PSO) with Genetic algorithm (GA) in computational grids. They only consider makespan and do not take into account trustworthiness of the network.

Ding et al. [30] proposed a reputation model for trust management in P2P grid environments. A context-related reputation function is proposed to promote an effective strategy for service provider selection, and simulation experiments demonstrate that important performance benefits can be obtained using this model.

Kouhei Umezaki et al. [31] proposed a Fuzzy-based Trustworthiness System for P2P overlays which uses JXTA protocol for nodes communications. This paper use Fuzzy Logic with these two parameters; Reputation and "Actual Behavior Criterion" to evaluate the Peer reliability. This paper uses the past experiences of the peers to detect the most reliable peer.

## 4. The Proposed Model

In our model, nodes are clustered into different groups based on the semantic similarity between their resources. The goal is to influence the semantic familiarity of each node to extract the location the node will inhabit in the P2P overlay.

Nodes are grouped together in the network space based on their character description in the semantic space. Suppose that A and B are two nodes with ontology sets S(A) and S(B), respectively. The similarity sim, between A and B, is the semantic concepts belonging to OSS of A and B: $|S(A) \cap S(B)|$. $\alpha$ and $\beta$ are parameters that depend on the application and how it wants to signify the differences between the two sets. These two parameters are the relative weights of the two nodes (A and B).

The semantic similarity between node A and node B is determined by equation (1), which is from [32]:

$$Sim(A,B) = \frac{|S(A) \cap S(B)|}{|S(A) \cap S(B)| + |\alpha(A) \cap S(B)| + |\beta(A) \cap S(B)|} \quad (1)$$

To address the necessity for frankness and the capability to be extended, we use the Web Ontology Language (OWL) [33], recommended by the W3C, as the ontology language. We use Chord [34], which is a distributed hash table (DHT) mechanism, to carry out distributed queries in the P2P environment. Inside any VO, the nodes are arranged as a P2P system. Reputation data are stored and collected by using super clusters to calculate global reputation scores. Each VO has a special node called the coordinator that is the super cluster of that VO. There is a trust agent in each super cluster which is responsible for storing reputation queries from nodes in other VOs and calculating the trust score of each node within the VO. The trust agent carries out trust management with the aid of fuzzy logic. The coordinator is used for communication between different VOs. Figure 1 shows the communication between different VOs. In this figure, P2, P5 and P10 are the coordinators inside of which the trust agents are located.

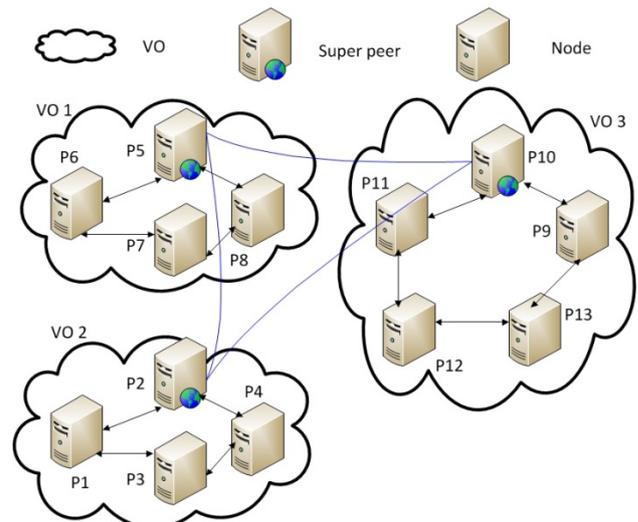

Fig. 1. Communication between different VOs

Regularly, a node will request a service from another node in the P2P grid environment. After a transaction between them occurs, the source node of the link will determine a score according to its evaluation of the service of the destination node of the link, and send it to the coordinator of the group. For example, after receiving some files from node P4, nodes P1, P2 and P3 assign the scores 0.2, 0.8 and 0.5, and send their scores to the coordinator of the group, which is P5. Figure 2 illustrates this. The red lines mean P1, P2 and P3 obtain services from P4, and the blue lines mean they send their scores of the quality of P4's services to P5, the coordinator of group. A trust agent is located in P5 which calculates the trust level of node P4. Peers who receive the service from a particular peer act as judges for the peer in question, and they then report reputation scores to the super peer. We call a node is malicious which its score is under the super peer threshold score.

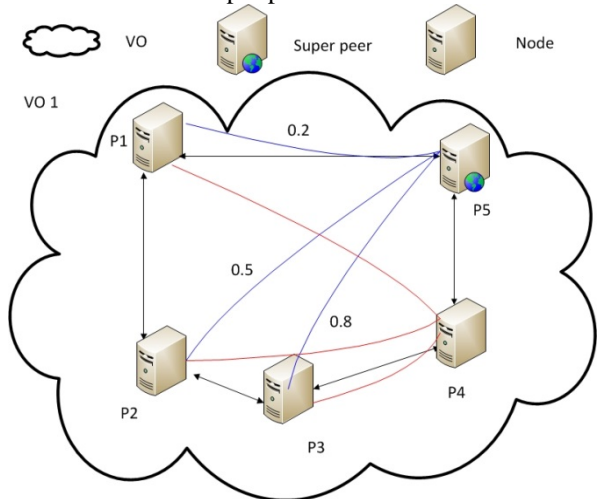

Fig. 2. Our model's trust management routine

Our model is based on fuzzy logic, and receives some scores as input parameters and uses fuzzy reasoning and a fuzzy inference system to generate a score as the system's output value, which is the node's trust level. Figure 3 illustrates this.

There are two common types of fuzzy inference systems: Mamdani and Sugeno. We used the Mamdani inference system because it is easy. Mamdani FIS has a major usage for capturing expert knowledge. It gives the feasibilities to describe the expertise in a good intuitive and human-like manner. Fuzzy inference consists of five stages: fuzzification of the input variables, carrying out the fuzzy operations and fuzzy reasoning, carrying out the fuzzy inferences, aggregation of the fired rules and defuzzification of the results. Each score, a crisp value, is represented by linguistic variables in fuzzy logic, and in the first step, fuzzification, a membership function is required to convert the scores achieved by a node into the fuzzy linguistic values Low, Medium or High, as shown in Figure 4. Each node which provides service is assessed based on the following membership function: choose one of the values Low, Medium, or High for the quality of services the node provides.

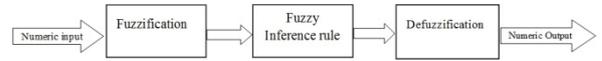

Fig. 3. The fuzzy inference system used

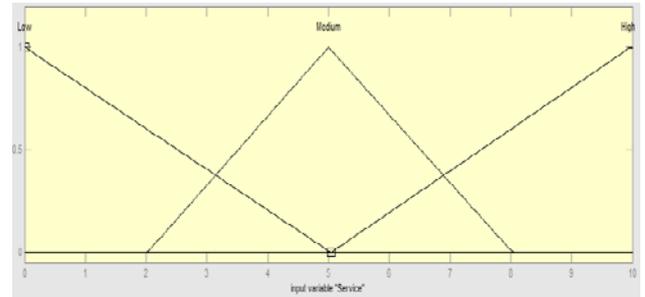

Fig. 4. Membership function for the quality of services

The level of trust can be calculated by using a fuzzy rule-based technique in the trust agent. The trust can be defined in terms of three levels: Low (L), Medium (M), and High (H). Some of the rules are shown in Table 1.

Fuzzy rules and fuzzy reasoning is the base fundamental of Mamdani fuzzy inference system. The applied fuzzy rules in this proposed system are gathered based on the human experiences and the proposed model hypothesizes. We note that, the suggested rules are able to be used in various Grid Environment based on grid administrator policies and account for input fuzzy system parameters.

The two input values are mapped to their separate membership degrees on their membership graphs. These degrees are compared and the least of the two values is then plotted onto the membership function in the output graph. The output graph represents the aggregation of the fired rules. After the output graph is created, the fuzzy output can be defuzzified into a crisp or numeric value, and the aggregation of all fired rules can be calculated as shown in Figure 5. We used the centroid method [35] to defuzzify the output as shown in equation (2):

$$\alpha = \frac{\int_Z \mu_A(x) z \, dz}{\int_Z \mu_A(x) \, dz} \quad (2)$$

where $\mu_A(x)$ is the output membership function of the system. Defuzzification means transforming the fuzzy plot (MF) to one scalar number. Therefore, we try to elicit a scalar number that present fuzzy set. The normal Defuzzification method is centroid method; this calculates the center point in the figure (plot). Specifically, center point is the average weight of each point in the domain. The weight of each node is corresponded the membership degree of the node.

The output indicates the trust degree of each node. A module called the surface viewer is used to exhibit the dependence of one of the outputs on any one or two of the inputs. The surface viewer is used for watching the entire output surface of the system, that is, the entire distance of the output set according to the entire distance of the input set. The surface viewer for our model is shown in Figure 6. This picture shows the output surface of the system based on the input value of two nodes.

Our proposed model architecture is comprised of three layers: the physical layer, the VO layer and the application layer, as shown in Figure 7. The lowest layer is the physical layer, which contains the available resources in our model. Each node has a local resource manager whose job is to manage local access to the resource. Resources usually include physical things such as computers and networks and comprise physical and logical resources. The next layer is the VO layer.

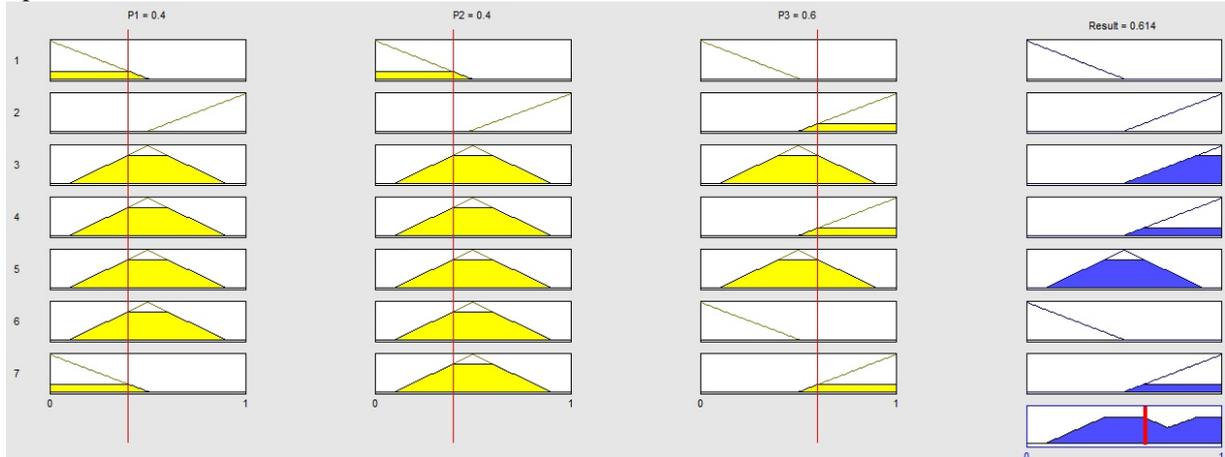

Fig. 5. The aggregated outputs

Table 1
Example of fuzzy rules

| P1 | P2 | P3 | Output |
|----|----|----|--------|
| L | L | L | L |
| H | H | H | H |
| L | M | M | M |
| H | H | M | H |
| M | M | M | M |
| L | M | H | M |
| L | M | L | L |
| H | M | H | H |
| L | L | H | L |
| M | M | H | H |

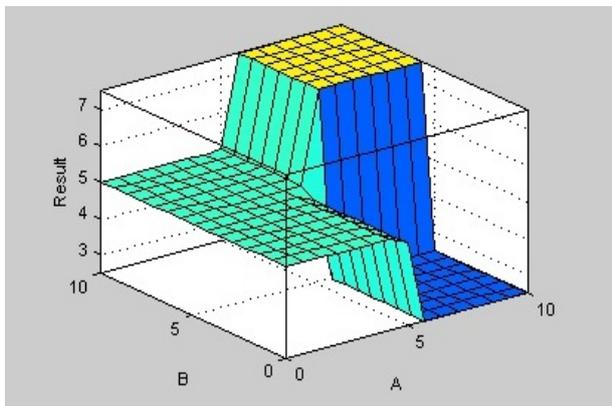

Fig. 6. The proposed model's surface viewer

The grid infrastructure is specified to represent the sharing and cooperation of diverse and distributed resources in VOs. We separate the VO layer into a collaboration services layer and a distributed resource coupling services layer using a P2P-based hybrid grid approach. Then trust management and searching for resources are divided in a suitable manner in order to integrate the distributed resources.

The distributed resources coupling services layer can efficiently manage the resources. Using P2P features in this layer, many systems can be developed to have fault-tolerance and good quality of service (QoS) such as P2P networks in [36]. The major duties of the cooperation services layer are trust management and delivering resources based on information received from the underlying layer. Nodes work with each other directly, without any centralized schedulers. Another role of this layer is to present ontological support for middleware software. OWL-DL [37] is used to achieve semantic goals. In typical Description logic (DL) systems, knowledge is divided into two components: the taxonomical box (T-Box) and the Assertional box (A-Box).

The T-Box collects conceptual knowledge about an entity and can be compared to the schema of a database. The A-Box provides the tangible knowledge about distinct entities within the domain. It is composed of concept assertions and role assertions. In the top layer, there are grid applications such as e-business and scientific computations and grid portals. One of the grid's tasks is to connect users and resources. Users use the grid portal to view and use grid resources; in this way the portal acts as a broker between users and resources. RDF [38] stands for "Resource Description Framework" which is used for describing the semantic similarity between resources. The following RDF triples are defined using some semantic relations in the semantic P2P grid ontology:

1. Node → Has a → Resource
2. Resource → Is a → File
3. File → Is a → Image

The following diagram shows the sematic relations of the P2P grid ontology.

```
<Rdfs:class    Rdf:ID="Node"/>

<Rdfs:class    Rdf:ID="resource">
   <Rdfs:SubClassOff Rdf:Resource="#Node"/>
</Rdfs:class>

<Rdfs:class    Rdf:ID="File">
 <Rdfs:SubClassOff Rdf:Resource="#Resource"/>
</Rdfs:class>

<Rdfs:class    Rdf:ID="Image">
   <Rdfs:SubClassOff Rdf:Resource="#File"/>
</Rdfs:class>
 </rdf:RDF>
```

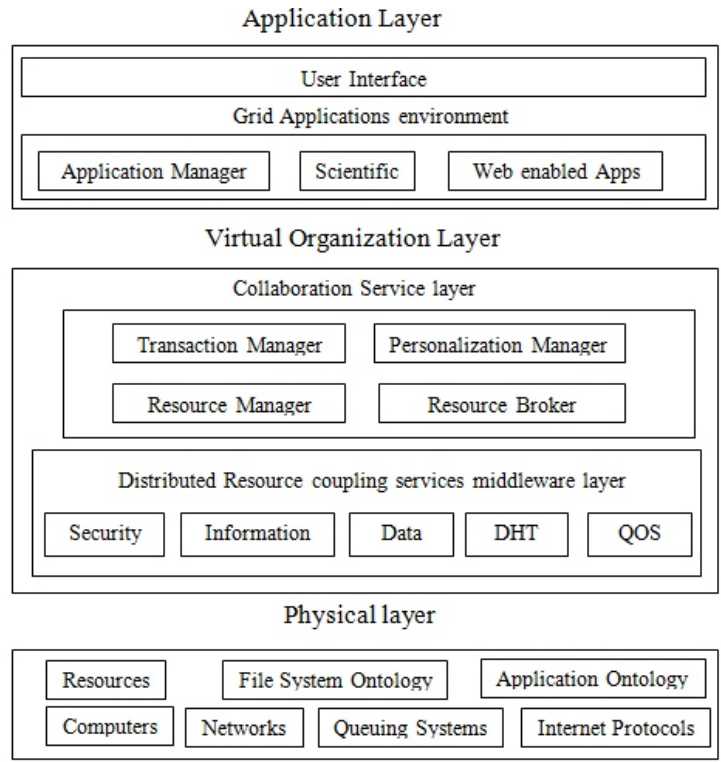

Fig. 7. The proposed model's architecture

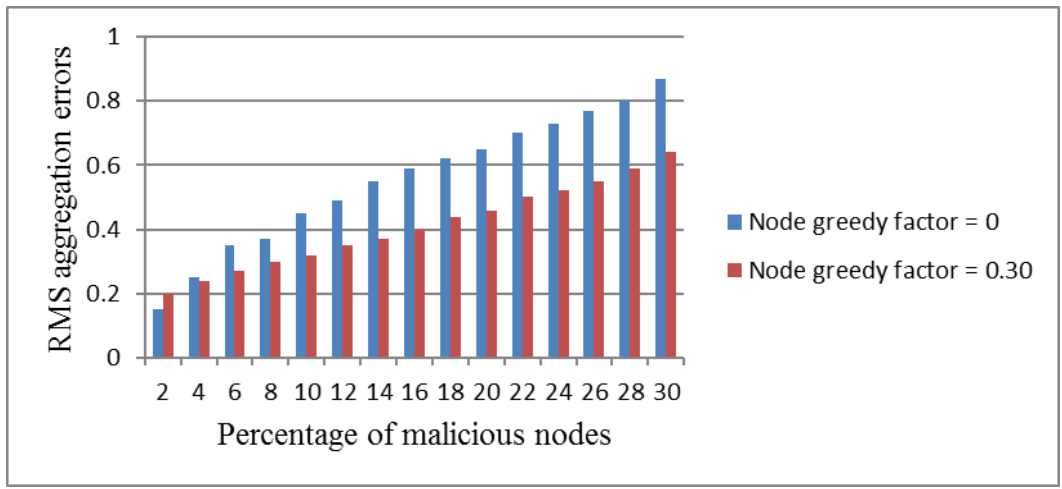

Fig. 8. RMS error

## 5. Performance Evaluation

To test the performance of our model, the two simulation experiments described below were carried out in some metrics such as root-mean-square (RMS) error and precision of malicious node detection using Matlab software [39] and Planetsim [40].

We compute below the RMS aggregation error in global scores with different percentages of malicious nodes in a semantic P2P grid environment. A lower RMS error indicates that the system is more robust to attacks by malicious nodes. The RMS error is determined by equation (3), which is defined in Power-Trust [23]:

$$RMS = \sqrt{\frac{\sum_{i=1}^{N}((v_i-u_i/v_i))^2}{N}} \quad (3)$$

where $v_i$ and $u_i$ are the computed and gossiped reputation scores of node $i$, respectively. We plot in Figure 8 the RMS aggregation errors under different values of α, the greedy factor that shows the desire of a node to work with chosen power nodes [41], and different percentages of malicious nodes. In this figure Alpha defines the percentage of malicious nodes and greedy factor is the eagerness for a node to work with other nodes.

For the next experiment we define a scenario and compare our approach with a fuzzy reputation based model which has usage in P2P overlays [38]. According to this scenario, there are 3 parameters; "neighbors' node scores" which is obtained from the nodes which get service form the considered node, Reputation (*R*) and Actual Behavioral Criterion (*ABC*). For our approach we consider the nodes scores as the input parameters of the fuzzy system; and we use R and ABC as the input parameters of fuzzy system for [38]. The fuzzy set values for the input parameters are between 0 and 1. Table 2 shows the result of our simulation.

Table 2

| ABC | R | P1 | P2 | P3 | Our PR | [31, 38] PR |
|---|---|---|---|---|---|---|
| 0.1 | 0.5 | 0.1 | 0.5 | 0.9 | 0.5 | 0.176 |
| 0.2 | 0.5 | 0.2 | 0.5 | 0.9 | 0.544 | 0.334 |
| 0.3 | 0.5 | 0.3 | 0.5 | 0.9 | 0.613 | 0.418 |
| 0.4 | 0.5 | 0.4 | 0.5 | 0.9 | 0.705 | 0.468 |
| 0.5 | 0.5 | 0.5 | 0.5 | 0.9 | ~0.8 | 0.491 |
| 0.6 | 0.5 | 0.6 | 0.5 | 0.9 | ~0.8 | ~0.5 |
| 0.7 | 0.5 | 0.7 | 0.5 | 0.9 | ~0.8 | ~0.5 |
| 0.8 | 0.5 | 0.8 | 0.5 | 0.9 | ~0.8 | ~0.5 |
| 0.96 | 0.5 | 0.96 | 0.5 | 0.9 | ~0.8 | ~0.5 |
| 1 | 0.5 | 1 | 0.5 | 0.9 | ~0.8 | ~0.5 |

In the rest of the evaluation of our model, we compare it with typical reputation methods [30, 42-44]. In this experiment, there are 100 nodes. As mentioned earlier, in a P2P environment, some peers may provide services with low quality; so a peer that may not be functioning may provide incorrect information that can deceive the whole network. It is therefore important in P2P grid networks to detect malicious peers. Since we use fuzzy logic in our model, we can achieve good accuracy. Our plan can detect malicious nodes with more precision. As Figure 9 illustrates, our model can detect more malicious nodes than a reputation model based on a trust cluster [42]. These results are due to our model's utilization of fuzzy theory. Computational complexity is a mathematical model for establishing reasonable proofs for algorithms. It studies the exact inherent difficulty of computational problems [45]. In our proposed fuzzy trust management model, all the rules in the rule base are processed in a parallel manner by the fuzzy inference engine. The search is thus performed in a parallel manner [19, 46], so the computational complexity is O (1).

In [30, 43], which use Chord as the search mechanism, if we consider the number of messages used to carry out trust management, the upper bound is given by O(logN), where N is the number of peers. In Chord, each peer maintains O(logN) neighbors. Figure 10 illustrates this, based on the number of nodes.

In [30, 44], graph nodes stand for peers, and after a transaction occurs between nodes, the source node of the link issues a feedback score according to its evaluation of the service of the destination node of the link. The edge label stands for the local trust score between the source and destination nodes. Consider Figure 2 again. In this example, the VO has 5 nodes.

Nodes P1, P2 and P3 download files from node P4. The global reputation is aggregated from all of P1, P2 and P3's local trust scores. The global reputation score of P4 is computed by weighting the three local scores. According to [30, 44], the global reputation for node P4 is calculated by equation (4):

$$R_4 = 0.2 * R_1 + 0.5 * R_2 + 0.8 * R_3 \quad (4)$$

Let's define $X = 0.2 * R_1$ and $Y = 0.5 * R_2 + 0.8 * R_3$, so we have $R_4 = X+Y$. Consider a situation where an error occurred in calculating X. The value of X will be between 0 and 1. If the real value of Y is 0.2, the value $R_4$ will be in the range between 0.2 and 1.2, as shown in Table 3. So the error has a significant effect on the result. Since our model uses fuzzy rules in which fuzzy sets overlap with each other, they are fired according to the inputs, and so the incorrect value of X has little effect on the output. We use 0.2, 0.5 and 0.8 as input parameters to the fuzzy system. We have computed the result of this error in our model with the aid of the Matlab software fuzzy toolbox [39].

Table 3

Comparison of models based on fault tolerance

| | Schemas | | |
|---|---|---|---|
| X value | FR Trust | Ding method [30] | Zhou method [44] |
| 0 | 0.2 | 0.2 | 0.601 |
| 0.1 | 0.3 | 0.3 | 0.601 |
| 0.2 | 0.4 | 0.4 | 0.617 |
| 0.3 | 0.5 | 0.5 | 0.647 |

| Schemas | | | |
|---|---|---|---|
| X value | FR Trust | Ding method [30] | Zhou method [44] |
| 0.4 | 0.6 | 0.6 | 0.634 |
| 0.5 | 0.7 | 0.7 | 0.722 |
| 0.6 | 0.8 | 0.8 | 0.783 |
| 0.7 | 0.9 | 0.9 | 0.791 |
| 0.8 | 1 | 1 | 0.791 |
| 0.9 | 1.1 | 1.1 | 0.791 |
| 1 | 1.2 | 1.2 | 0.791 |

Table 2 shows the effect of this error on our model and on the other methods [30, 44]. We should mention that in our model, we do not have an output greater than 1. The aim of this comparison is to evaluate our model in terms of fault tolerance. In our method, the results of each situation (the different values of X between 0 and 1) are very similar to one another and are in the same set. This is because of the overlap between the rules.

The coverage feature in the method adds fault tolerance to the method. Refer to Table 2: because of the coverage in the fuzzy rules, when there is a failure in the x value, this has a negligible impact in our method, with fewer effects on the final results.

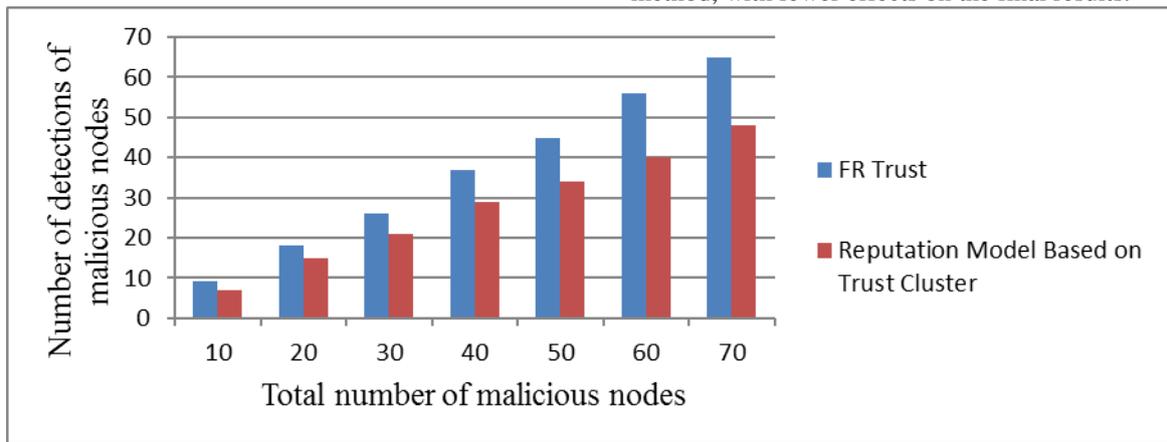

Fig. 9. Compression of accuracy

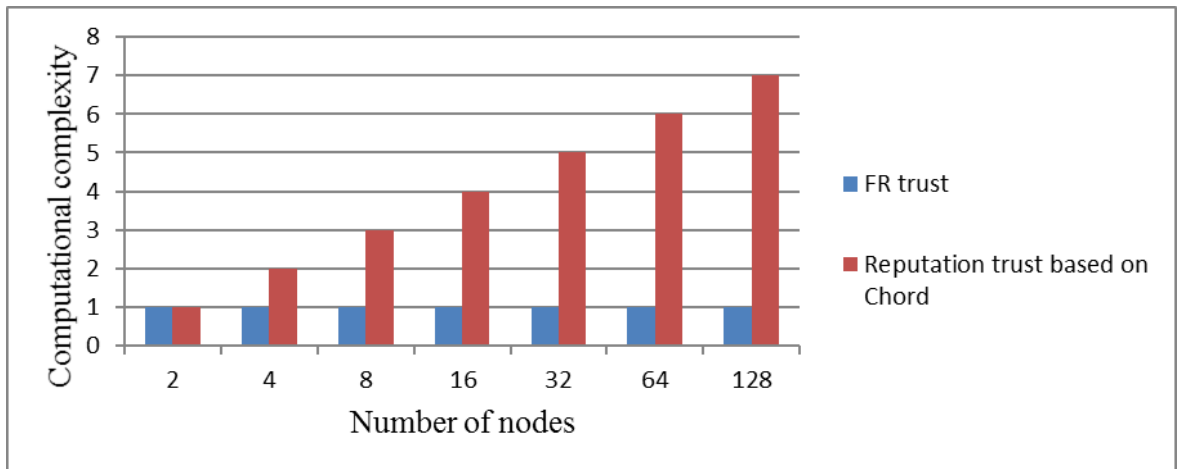

Fig. 10. Computational complexity

## 6. Conclusion and future work

In any P2P environment, trust management is quite expensive when the network grows to reach millions of nodes. In a P2P reputation model, speed and accuracy are important. An additional important factor for evaluating the model is whether it is robust against malicious nodes. This paper presents a trust management model based on fuzzy theory in a semantic P2P grid environment. Our model integrates two well-known approaches, fuzzy theory and a reputation model, to gather locally-created feedback and produce a global node trust degree. Our model makes important performance gains in speed and accuracy and is robust to malicious nodes, with low computational complexity.

The outlook of this research is the employment of semantic as a fuzzy parameter for computing a peer trust level. As we mentioned earlier, at this paper, the semantic is just used for clustering the grid environment based on the similarity between resources. We expect that semantic as a fuzzy parameter increases peer trust level. We plan to extend this research and apply it to our approach as our future work.

**Acknowledgment**

The authors of this paper would like to thank Mr ebrahim dashti rahmat abadi and Miss Safiye Ghasemi (Ph.D. students at the Islamic Azad University- Tehran Science and Research Branch) and Miroslaw Korzeniowski from Wroclaw University of Technology and Mr Damià Castellà (A researcher & research fellow at the University of Lleida) for their kindly comments and advices.